\documentclass{article}

\usepackage{arxiv}
\usepackage{authblk}
\usepackage{subfig}
\usepackage{graphicx}
\usepackage{epstopdf, epsfig}

\usepackage[export]{adjustbox}
\usepackage{amsmath,amssymb}
\usepackage{booktabs}
\usepackage{longtable,tabularx}
\usepackage{xcolor}
\usepackage{cleveref}
\usepackage{makecell}
\usepackage{tablefootnote} 

\usepackage{url}

\crefformat{section}{\S#2#1#3}

\newenvironment{acknowledgement}{%
  \section*{Acknowledgement} 
}{}

\newcommand{\architecture}{ZipGAN}
\newcommand{\cnn}{ZipCNN}

\newcommand{\newpart}[1]{{\color{magenta}}}

\newcommand{\DNSsub}{\mathrm{{orig}}}

\newcommand{\ENCsub}{\mathrm{{enc}}}
\newcommand{\SRsub}{\mathrm{{dec}}}

\newcommand{\grad}{\nabla}

\title{\architecture: Super-resolution-based generative adversarial network framework for data compression of direct numerical simulations}

\author[1]{L. Nista}
\author[2]{C. D. K. Schumann}
\author[1]{F. Fr\"{o}de}
\author[1]{M. Gowely}
\author[3]{T. Grenga}
\author[4]{J. F. MacArt}
\author[5]{A. Attili}
\author[1]{H. Pitsch}

\affil[1]{Institute for Combustion Technology, RWTH Aachen University, Aachen, 52056, Germany}
\affil[2]{Department of Engineering, University of Cambridge, Cambridge, CB2 1PZ, United Kingdom}
\affil[3]{Department of Aeronautics and Astronautics, Faculty of Engineering and Physical Sciences, University of Southampton, Southampton, SO17 1BJ, United Kingdom}
\affil[4]{Aerospace and Mechanical Engineering, University of Notre Dame, Notre Dame, IN 46556, USA}
\affil[5]{School of Engineering, Institute for Multiscale Thermofluids, University of Edinburgh, Edinburgh, EH93FD, United Kingdom}

\begin{document}

\maketitle

\begin{abstract}
The advancement of high-performance computing has enabled the generation of large direct numerical simulation (DNS) datasets of turbulent flows, driving the need for efficient compression/decompression techniques that reduce storage demands while maintaining fidelity. Traditional methods, such as the discrete wavelet transform (DWT), cannot achieve compression ratios of 8 or higher for complex turbulent flows without introducing significant encoding/decoding errors. On the other hand, a super-resolution-based generative adversarial network (SR-GAN), called \architecture, can accurately reconstruct fine-scale features, preserving velocity gradients and structural details, even at a compression ratio of 512, thanks to the more efficient representation of the data in a compact latent space. Additional benefits are ascribed to adversarial training. The high GAN training time is significantly reduced with a progressive transfer learning approach and, once trained, they can be applied independently of the Reynolds number. It is demonstrated that \architecture\ can enhance dataset temporal resolution without additional simulation overhead by generating high-quality intermediate fields from compressed snapshots. The \architecture\ discriminator can reliably evaluate the quality of decoded fields, ensuring fidelity even in the absence of original DNS fields. Hence, \architecture\ compression/decompression method presents a highly efficient and scalable alternative for large-scale DNS storage and transfer, offering substantial advantages over the DWT methods in terms of compression efficiency, reconstruction fidelity, and temporal resolution enhancement.
\end{abstract}

\section{Introduction}
\label{sec:introduction}

The advancement of high-performance computing and modern cluster architectures, along with cutting-edge experimental facilities, has enabled the generation of vast high-fidelity datasets from direct numerical simulations (DNSs) and experiments. As simulations and experiments increase in fidelity, the challenges related to the storage and transfer of large three-dimensional spatiotemporal data have become more prominent \cite{hilbert2016big}. For instance, numerical simulations utilizing thousands of processor cores can produce three-dimensional datasets that span several terabytes in size \cite{ihme2022combustion}. Nevertheless, as DNS and experimental data must be stored for post-processing analysis, data transfer, long-term archiving, and adherence to FAIR principles (Findability, Accessibility, Interoperability, and Reusability), there is a pressing need for effective and scalable data compression techniques. These techniques aim to minimize the storage requirements of datasets while preserving the accuracy required for meaningful analysis.

\emph{In situ} data compression is commonly employed to minimize storage requirements. This compresses (encodes) the original, uncompressed data, $\mathrm{\phi_{orig}}(\mathbf{x},t) \in \mathbb{R}^m$,  into a reduced representation, $\mathrm{\phi_{enc}}(\mathbf{x},t) \in \mathbb{R}^n$, such that $n \ll m$. This compression process reduces file size by eliminating spatiotemporal statistical redundancies. Generally, the compression process is handled by an encoder $\Lambda$, while the decompression process is handled by a decoder $\Gamma$, such that
\begin{equation}
    \mathrm{\phi_{enc}}(\mathbf{x},t) = \Lambda(\mathrm{\phi_{orig}}(\mathbf{x},t)), \, \, \mathrm{\phi_{orig}}(\mathbf{x},t) \approx \mathrm{\phi_{dec}}(\mathbf{x},t) = \Gamma(\mathrm{\phi_{enc}}(\mathbf{x},t)),
    \label{eqn:theory_compression_data}
\end{equation}
where the decompressed (decoded) dataset $\mathrm{\phi_{dec}}(\mathbf{x},t)$ is identical to $\mathrm{\phi_{orig}}(\mathbf{x},t)$ in the case of lossless compression.
Conversely, encoding/decoding errors are inherent to lossy compression, and the nature of the errors depends on the compression ratio, data structure, and the encoder and decoder algorithms.

Lossless compression techniques, such as SZIP\footnote{\url{https://support.hdfgroup.org}}, are commonly applied to reduce the size of flow field data, as these methods ensure that the original data can be reconstructed without spurious artifacts. However, the compression ratio achieved by lossless techniques is typically low---usually less than 2---which is often insufficient to compress large-scale flow datasets \cite{7322799}.

Common lossy compression techniques for flow field data include modal analysis \cite{taira2017modal}, sub-sampling, local re-simulation \cite{wu2020data}, and wavelet-based methods \cite{schneider2010wavelet}. These techniques offer significantly higher compression ratios of 4 or more, depending on the application. Due to their significantly higher compression, we consider only lossy compression methods in the remainder of this section.

Popular image compression algorithms such as the JPEG2000 standard have been evaluated for their suitability in DNS data compression \cite{schmalzl2003using}. Similarly, Anatharaman \textit{et al.}~\cite{anatharaman2023image} explored the compression of spatiotemporal fluid-flow data using multimedia techniques. They found that standard video compression methods, such as AVI and H.265, performed well in certain cases, and proper orthogonal decomposition (POD) also contributed to effective compression. Decomposition methods such as POD \cite{berkooz1993proper} and dynamic mode decomposition (DMD) \cite{schmid2010dynamic} have also been employed on their own as data compression techniques.

Wavelet methods have been shown to offer efficient compressed representations of turbulent flows \cite{farge1992wavelet, grenga2018dynamic}, and several lossy compression techniques using wavelet-based approaches have been investigated \cite{schneider2010wavelet, kolomenskiy2022waverange}. While these techniques enable high parallel performance for large datasets, they often face difficulties in balancing the trade-off between compression ratio and decoding accuracy. This is the case particularly in complex, high-dimensional datasets such as those generated by homogeneous isotropic turbulence (HIT) DNSs \cite{10.2514/6.2023-1684}.

Recent advances in deep learning (DL) have opened new avenues for data compression, promising higher compression ratios while maintaining data integrity \cite{taira2020modal}. This is motivated by the ability of DL architectures to efficiently capture nonlinearities inaccessible to traditional methods, which enables DL techniques to create more compact and efficient compressed representations. Furthermore, the information is compressed to a low-dimensional latent space, in which the model can represent essential features and patterns more effectively. This approach allows for the retention of critical data characteristics while reducing redundancy, leading to significant improvements in both compression efficiency and the quality of the decompressed output.

Glaws \textit{et al.}~\cite{PhysRevFluids.5.114602} proposed a fully convolutional autoencoder and showed that for fixed accuracy (decompression error), which was obtained by applying standard singular value decomposition methods (such as POD or DMD), a substantially higher compression ratio of up to 64 can be achieved. Similarly, Momenifar \textit{et al.}~\cite{doi:10.1080/14685248.2022.2060508} developed a physics-informed DL architecture based on vector quantization, producing a discrete low-dimensional representation of data from three-dimensional turbulent flow simulations. Their model achieved a compression ratio of 85 and accurately reproduced large-scale flow statistics with nearly identical decompressed fields. However, deviations in the decompressed fields were found to occur at the smallest scales.

Data compression can also be obtained through DL super-resolution (SR) approaches, which aim to accurately reconstruct data from its compressed (low resolution---LR) form. Guo \textit{et al.}~\cite{9086293} proposed a unified SR-convolutional neural network (CNN) architecture comprising three separate CNNs. Each CNN processes one component of the encoded vector field, and together they produce the synthesized decompressed vector field. Their approach, tested on various three-dimensional datasets, achieved compression ratios ranging from 64 to 512. Additionally, Gao \textit{et al.}~\cite{10.1063/5.0054312} trained an SR-CNN using only encoded fields, without the need for their original counterparts, and obtained a compression ratio of 400 for two-dimensional flow. This approach leverages physics and boundary condition constraints to enhance compression efficiency. Matsuo \textit{et al.}~\cite{matsuo2024reconstructing} proposed a method combining a supervised CNN with adaptive sampling-based super-resolution analysis to improve data compressibility, achieving a compression ratio of around 1000. However, this method was only applied to a simple cylinder wake dataset, and comparable compression ratios have yet to be achieved for complex turbulent flows. Sofos \textit{et al.}~\cite{SOFOS2024106396} compared five different SR-CNNs, reporting that U-Net-based models with skip connections produced sharper features in regions prone to blurring errors. Similar results were observed in short-range forecasts of near-surface winds \cite{https://doi.org/10.1002/met.1961}.

SR-CNNs have shown promise in reconstructing compressed data, particularly for relatively simple flow structures \cite{fukami2023super}. However, SR-generative adversarial networks (GANs) may offer additional advantages in data decompression due to their ability to accurately reconstruct high-frequency and small-scale features that are often blurred by CNN-based models, as recently demonstrated by Nista \textit{et al.}~\cite{nista2024PRF} in the context of super-resolution reconstruction. This capability might make SR-GANs well-suited for improving compression ratios without compromising the fidelity of the decompressed data, in particular for turbulent flows. At the same time, SR-GAN-based compression techniques have achieved higher compression ratios compared to both traditional methods and SR-CNN-based approaches while maintaining fidelity to the original data for image-processing applications \cite{tolunay2018generative}. Notably, Agustsson \textit{et al.}~\cite{9010721} jointly trained a GAN composed of an encoder, a decoder/generator, and a multiscale discriminator. Their approach produced visually appealing decompressed images even at compression ratios up to 16, at which state-of-the-art methods typically begin to introduce noticeable artifacts. Nonetheless, there is a paucity of studies on the use of SR-GANs for data compression of complex, three-dimensional turbulent flow datasets.

This study explores the potential of a SR-based GAN, called \architecture, for compressing and decompressing three-dimensional HIT DNS datasets for storage and transfer. The performance of \architecture\ as a data decompression method is compared with the widely used discrete wavelet transform method, and the maximum compression ratio at which the accuracy of the decompressed field begins to degrade is determined. Additionally, the computational costs of both the encoding and decoding processes are evaluated for each method, with special attention devoted to the training cost of the \architecture. Furthermore, unique techniques that leverage the capabilities of the \architecture\ framework are explored.


\section{Datasets and data compression methods}
\label{sec:dataset_and_preprocessing}

Two forced HIT DNS datasets are considered to investigate the effectiveness of the \architecture\ framework (introduced in Sec.~\ref{sec:architecture}) for the compression of complex turbulent flows. The Taylor-microscale Reynolds numbers of the datasets are $\mathrm{Re_{\lambda}} \approx 90$ (Re90) and $\mathrm{Re_{\lambda}} \approx 210$ (Re210). The domain is discretized on 256 $\times$ 256 $\times$ 256 points for Re90 and 1024 $\times$ 1024 $\times$ 1024 points for Re210. Linear forcing $\textbf{f}= \textit{A} \textbf{u}$ is applied in both simulations, where $A$ is a forcing parameter that is inversely proportional to the eddy turnover time. Simulations are initialized with a synthetic von Kármán--Pao spectrum \cite{Nista_datapublication}. The minimum cell size, $\mathrm{dx}$, satisfies $\mathrm{\kappa_\texttt{max} \eta \geq 1}$, where $\mathrm{\eta}$ is the Kolmogorov length scale and $\mathrm{\kappa_\texttt{max}}$ is the maximum discrete wavenumber. The simulation parameters are reported in Table \ref{tab:simulation_parameters}. Only the Re90 dataset is used for training the \architecture, while the Re210 dataset is reserved for the generalization tests conducted in Sec.~\ref{sec:results}. Additional details of these datasets can be found in previous works \cite{nista2024PRF, Nista_datapublication}. 
\begin{table}[!ht]
\begin{center}
\def~{\hphantom{0}}
\begin{tabular}{ccccccccc}
\toprule
Case & $\mathrm{N}$ & $\mathrm{Re_{\lambda}}$  & $\mathrm{Re_t}$  & $\mathrm{dx/\eta}$ & $\mathrm{\# \, snapshots}$ & $\mathrm{S_{case}}$ [GB] & Train & Test \\
\midrule
Re90  & $256^3$  & 90 & 455 & 1.98 & 250 & 93 & $\bullet$ & $\bullet$ \\
Re210 & $1024^3$ & 210 & 1300 & 1.97 & 130 & 3120 & & $\bullet$ \\
\bottomrule
\end{tabular}
\caption{Simulation parameters for training and testing datasets. $\mathrm{N}$ is the number of mesh points on each spatial dimension, $\mathrm{Re_t}$ is the turbulent Reynolds number, $\mathrm{dx/\eta}$ is the mesh resolution relative to the Kolmogorov length scale $\mathrm{\eta}$, $\mathrm{\# \, snapshots}$ is the total number of snapshots extracted from each dataset, and $\mathrm{S_{case}}$ is the total dataset size in GB. ``Train'' and ``Test'' flags indicate datasets used for model training and testing.}
\label{tab:simulation_parameters}
\end{center}
\end{table}

After the Re90 simulation reached a statistically stationary state, 250 snapshots of the three-dimensional velocity vector, $\mathrm{\textbf{u}}$ = $(u, v, w)^{T}$ were considered, with five snapshots extracted per eddy-turnover time. To obtain encoded (compressed) fields $\mathrm{\phi_{enc}}$ of different compression ratios, all available snapshots of the instantaneous velocity were filtered using a box filter kernel with a discrete downsampling operation. Multiple kernel widths $\mathrm{\Delta = n_{\Delta}\,} \mathrm{dx}$ were used with downsampling factors $\mathrm{n_\Delta \in [2, 4, 8, 16]}$. This preprocessing step is denoted as an ``encoding operation.'' The training, validation, and testing datasets are composed of original DNS fields $\mathrm{\phi_{orig}}$ and the corresponding encoded fields $\mathrm{\phi_{enc}}$. The first 190 snapshots of the Re90 dataset were used for training, while the last 60 snapshots were used for validation and testing (training/validation ratio $\approx 0.75$/$0.25$). The validation was then performed on 20 in-sample fields not employed for training. The last 20 snapshots were used to obtain the averaged results discussed in Sec.~\ref{sec:results}. The time separation between training/validation samples and testing samples was more than 10 eddy-turnover times, thus, the two sets of samples are considered statistically uncorrelated.

Loading the entire training dataset into GPU memory is impractical given memory limitations. To alleviate this, a patch-to-patch training strategy is applied, where sub-domains are randomly extracted from the full domain for each snapshot \cite{Nista2024_parallel}. The \architecture\ framework (Sec.~\ref{sec:architecture}) was trained using $\mathrm{\phi_{enc}}$ and $\mathrm{\phi_{orig}}$ subboxes. Conversely, during the evaluation of testing samples, the entire domain is reconstructed continuously, without employing subboxes as used for training. The velocity components of the original and encoded fields used for training and testing were normalized by the global maximum and minimum of $\mathrm{\phi_{orig}}$ to improve the network's performance \cite{nista2023proci}. Training was performed on the J{\"u}lich DEEP-EST cluster (DEEP-DAM partition) using four nodes, each containing one NVIDIA V100 32GB GPU, using the TensorFlow framework \cite{tensorflow2015-whitepaper}. The Adam optimizer \cite{kingma2014adam} was utilized with initial learning rate $\alpha_0=10^{-4}$ for consistency with previous applications \cite{nista2022SciTech, grenga_CST}.


\subsection{Super resolution generative adversarial network architecture (\architecture) and training loss function for data decompression}
\label{sec:architecture}

\begin{figure}[bh] 
    \centering
    \includegraphics[width=\textwidth]{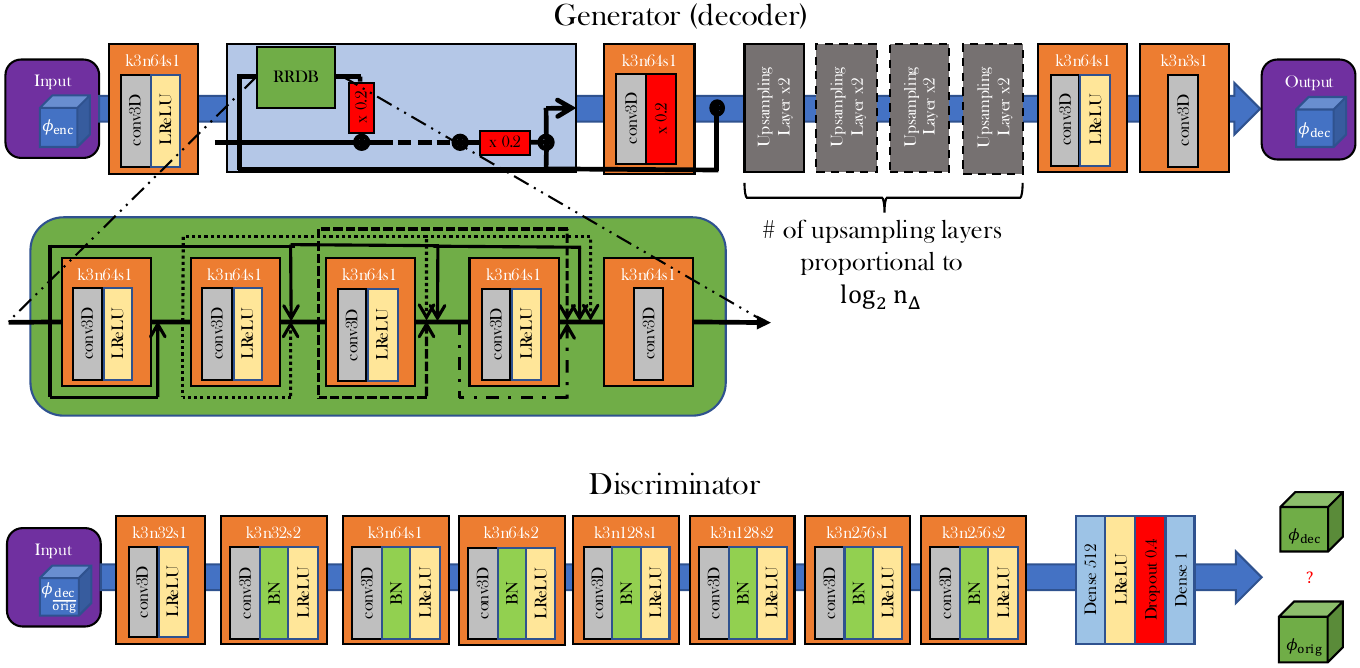}
    \caption{The generator (decoder) and discriminator structures employed by the \architecture\ architecture. Input and output fields reported in Eq.~\ref{eqn:theory_compression_data} are highlighted.
    Each convolutional block contains kernels of size $k$, $n$ filter maps, and $s$ strides along each spatial dimension of the convolutional layer. \cnn\ specifically refers to the generator of \architecture, which is trained in a fully supervised manner without the adversarial contribution present in the complete GAN framework.}
    \label{fig:TSRGAN_architecture}
\end{figure}

The \architecture\ framework shown in Fig.~\ref{fig:TSRGAN_architecture}, initially designed for small-scale turbulence reconstruction, is employed in this work as a decoder tool. The architecture, based on previous works in the context of turbulence modeling, has demonstrated superior reconstruction performance for both in-sample and out-of-sample HIT flow configurations compared to standard supervised CNN-based architectures \cite{nista2024PRF}. Its adoption in this study is motivated by its proven efficacy in previous works, while its optimization for data decompression is currently ongoing and beyond the scope of this work.

This architecture employs a conventional GAN setup for training, consisting of a generator and a discriminator. The generator, employed as a decoder, super-resolves the encoded field $\mathrm{\phi_{enc}}$ to generate the decoded field $\mathrm{\phi_{dec}}$. The discriminator distinguishes between $\mathrm{\phi_{dec}}$ and $\mathrm{\phi_{orig}}$ \cite{goodfellow2020generative}. As a result, both networks improve their predictions during the adversarial training process. The generator, a CNN based on the SRResNet architecture \cite{he2016deep}, is designed to recover small-scale features using skip connections and three-dimensional convolutional layers. A central feature of the generator is the residual-in-residual dense block (RRDB)  architecture, which includes residual connections and densely connected layer blocks. The generator also includes upsampling layers, which use nearest-neighbor interpolation followed by a convolution layer to enhance the operation. Each upsampling layer doubles the spatial resolution, hence, the number of upsampling layers is determined by $\Delta$. The total number of trainable parameters of the generator is approximately 10 million, and is mainly driven by the number of filter kernels per convolutional layer (a constant), and is nearly independent of the number of upsampling layers. The discriminator (Fig.~\ref{fig:TSRGAN_architecture}) is structured as a deep deconvolutional neural network with fully connected and convolutional layers, similar to the generator. It outputs a binary classification and contains around 12 million trainable parameters.

The generator's loss function ($\mathcal{L}_{\mathrm{GEN}}$) is the weighted sum of four loss components: pixel loss ($L_{\mathrm{{pixel}}}$), pixel gradient loss ($L_\mathrm{{gradient}}$), continuity loss ($L_\mathrm{{continuity}}$), and the adversarial loss ($L_\mathrm{{adversarial}}$). The discriminator's loss function ($\mathcal{L}_{\textsc{DISC}}$) is based on the relativistic average GAN loss function. These are given by
\begin{equation}
\begin{aligned}
     \mathcal{L}_{\textsc{GEN}} &=  \beta_1 \, L_{\mathrm{pixel}} + \beta_2 \, L_\mathrm{{gradient}} + \beta_3 \, L_\mathrm{{continuity}} + \beta_4 \, L_\mathrm{{adversarial}} \\
     L_\mathrm{{pixel}} &= \mathrm{MSE}(\phi_\SRsub, \phi_\DNSsub) \\
     L_\mathrm{{gradient}} &= \mathrm{MSE}(\grad \phi_\SRsub, \grad \phi_\DNSsub) \\
     L_\mathrm{{continuity}} &= \mathrm{MSE}(\nabla \cdot \phi_\SRsub, 0) \\
     L_\mathrm{{adversarial}} &= -\mathbb{E}_{\phi_{\ENCsub} \sim p_\mathrm{enc}}[\log(\sigma(D(G(\phi_\ENCsub)) - \mathbb{E}_{\phi_{\DNSsub} \sim p_\mathrm{orig}}[D(\phi_\DNSsub)]))]\ \\
    &\qquad - \mathbb{E}_{\phi_{\DNSsub} \sim p_\mathrm{orig}}[\log(1 - \sigma(D(\phi_\DNSsub) - \mathbb{E}_{\phi_{\ENCsub} \sim p_\mathrm{enc}}[D(G(\phi_\ENCsub))]))] \\
    \mathcal{L}_{\textsc{DISC}} =\ &\mathbb{E}_{\phi_{\DNSsub} \sim p_\mathrm{orig}}[\log(\sigma(D(\phi_\DNSsub) - \mathbb{E}_{\phi_{\ENCsub} \sim p_\mathrm{enc}}[D(G(\phi_\ENCsub))]))]\ \\
&+ \mathbb{E}_{\phi_{\ENCsub} \sim p_\mathrm{enc}}[\log(1 - \sigma(D(G(\phi_\ENCsub)) - \mathbb{E}_{\phi_{\DNSsub} \sim p_\mathrm{orig}}[D(\phi_\DNSsub)]))] \, ,
\label{eqn:loss_generator}
\end{aligned}
\end{equation}
where $\mathrm{MSE}(a,b)$ is the mean-squared error between $a$ and $b$, $\mathbb{E}[\cdot]$ is the mathematical expectation computed with either the encoded distribution samples, $p_\mathrm{enc}$, or the original distribution samples, $p_\mathrm{orig}$, $\sigma(\cdot)$ is the sigmoid function, and $D(\phi_\DNSsub)$ and $G(\phi_\ENCsub)$ are the outputs of the discriminator and generator, respectively. The hyperparameters $\beta_{i} = [0.89, 0.06, 0.05, 6 \cdot 10^{-5}]$ were selected to ensure that each term in $\mathcal{L}_{\textsc{GEN}}$ is of the same order. Further details on the choices of $\beta$ are provided in \cite{nista2024PRF}. 

Challenges associated with training GANs include convergence issues caused by parameter oscillations and training instability \cite{goodfellow2020generative}. To minimize their impact, the generator is first pretrained in a fully supervised manner using only $L_{\mathrm{pixel}}$ for downsampling factor $\mathrm{n}_{\Delta} = 2$. This supervised pretraining continues until the loss function and statistics of the reconstructed field stabilize, indicating convergence. The pretrained generator is then used as the starting point for the GAN training, ensuring that each training run is initialized with the same generator weights. The discriminator is not pretrained during this phase. Moreover, to assess the impact of adversarial training on data compression, \cnn\ is introduced, in which only the generator component of \architecture\ is utilized. The \cnn\ generator is trained independently in a fully supervised manner using  $L_{\mathrm{pixel}}$, $L_{\mathrm{gradient}}$, and  $L_{\mathrm{continuity}}$, without incorporating adversarial contributions. The purpose of this baseline is to isolate the effects of adversarial training, allowing for a direct comparison between adversarial and purely supervised training approaches.

Moreover, transfer learning (TL) is applied to \architecture\ for accelerating the training process for $\mathrm{n}_{\Delta} \in [4, 8, 16]$ as described in Sec.~\ref{sec:results}. In this training approach, models are progressively fine-tuned to increase upsampling factors. The converged model trained for $\mathrm{n}_{\Delta}=2$ is used to initialize training for $\mathrm{n}_{\Delta}=4$, which is used to initialize the model for $\mathrm{n}_{\Delta}=8$, which in turn is used to initialize the model for $\mathrm{n}_{\Delta}=16$. All model layers remain trainable, enabling full adaptation to each new task. From a network architecture perspective, only the upsampling layers of the generator (in both \cnn\ and \architecture\ frameworks) are adjusted, and fine-tuning is done with a reduced learning rate to ensure stable results.

\subsection{Discrete wavelet transform method for data compression}

We evaluate the performance of the \architecture\ framework for data compression against the discrete wavelet transform (DWT), a traditional wavelet-based technique. The DWT is well-suited for its efficiency in managing large datasets due to its lower computational complexity when compared to methods such as DMD \cite{schneider2010wavelet}.

In the DWT, a three-dimensional velocity field is decomposed into multiple frequency components through localized wavelet basis functions. This decomposition captures the spatial and temporal modes essential for accurately compressing complex flow structures.
The compressed fields are expressed in terms of wavelet coefficients $d_{j,k}$, scaling functions, $\Phi_{0,k}(x)$, and wavelet functions $\Psi_{j,k}(x)$, where $j$ is the resolution level ($j=0$ being the coarsest level) and $k$ is the translation index. The one-dimensional wavelet transform of a continuous function $f$ is given by 
\begin{equation}
\begin{aligned}
f_\mathrm{{orig}}(\mathbf{x},t) \approx f_\mathrm{{dec}}(\mathbf{x},t) & =  \sum_{k} f_{\mathrm{orig},k} \Phi_{0,k}(x) + \sum_{j = 0}^{J} \sum_{\{k \; : \; |d_{j,k}| \geq \varepsilon\}} d_{j,k} \Psi_{j,k}(x) \\
& {} \qquad +  \sum\limits_{j=0}^{J} \sum_{\{
k \; : \; |d_{j,k}| < \varepsilon\}} d_{j,k} \Psi_{j,k}(x).
\label{eqn:wavelet}
\end{aligned}
\end{equation}
The wavelet coefficients provide a direct measure of the local approximation error at each location. Thus, compression is realized by discarding any wavelet coefficients smaller than a thresholding parameter $\varepsilon$, which is determined in order to match the target compression ratio.

For this analysis, the Biorthogonal 4.4 (Bior4.4) wavelet is chosen due to its advantageous properties, including symmetry, compact support, and high reconstruction accuracy \cite{kolomenskiy2022waverange}. The significant coefficients are retained for the reconstruction of the original three-dimensional flow field through the inverse DWT. A detailed explanation of this approach can be found in \cite{kolomenskiy2022waverange}.


\section{Results}
\label{sec:results}

The performance of DWT and \architecture\ compression methods applied to the datasets is evaluated for various compression ratios, denoted $\theta$, and compared with the original DNS snapshots, as shown in Table \ref{tab:table_HIT_results}. For $\theta = 2$ and $\theta = 4$, the \architecture\ could not be applied due to the limitations of its three-dimensional upsampling layers, necessitating $\theta\geq 8$.
\begin{table}[!ht]
    \centering
    \begin{tabular}{cccccccc}
    \toprule
    \makecell{Compression \\ method} & $\mathrm{\theta}$ & $\mathrm{n_{\Delta}}$ & \multicolumn{2}{c}{\makecell{$\mathrm{S_{\phi_{enc}}}$ $[\mathrm{MB}]$}} & \multicolumn{2}{c}{\makecell{$\mathrm{N_{\phi_{enc}}}$}} & \makecell{\architecture\ $[\mathrm{MB}]$} \\
    \cmidrule(lr){4-5} \cmidrule(lr){6-7} 
     & & & Re90 & Re210 & Re90 & Re210 & \\
    \midrule 
    Original & 1 & -- & 380   & 24576 & $256^3$ & $1024^3$ & --  \\
    \hline
    DWT & 2 & -- & 190 & 12288 & -- & -- & --  \\
    \cline{1-8}
    DWT & 4 & -- & 95 & 6144 & -- & -- & --  \\
    \cline{1-8}
    \makecell{DWT \\ \architecture} & \makecell{8} & \makecell{-- \\ $2$} & 48 & 3072 & \makecell{-- \\ $128^3$} & \makecell{-- \\ $512^3$} & \makecell{-- \\ 42}  \\ 
    \cline{1-8}
    \makecell{DWT \\ \architecture} & \makecell{64}  & \makecell{-- \\ $4$}  & 6 & 410 & \makecell{-- \\ $64^3$}  & \makecell{-- \\ $256^3$} & \makecell{-- \\ 42}  \\
    \cline{1-8}
    \makecell{DWT \\ \architecture} & \makecell{512} & \makecell{-- \\ $8$}  & 0.8 & 51 & \makecell{-- \\ $32^3$}  & \makecell{-- \\ $128^3$} & \makecell{-- \\ 43}  \\
    \cline{1-8}
    \makecell{DWT \\ \architecture} & \makecell{4096} & \makecell{-- \\ $16$} & 0.09 & 6 & \makecell{-- \\ $16^3$}  & \makecell{-- \\ $64^3$} & \makecell{-- \\ 44}  \\
    \bottomrule
    \end{tabular}
    \caption{Comparison of the encoding/decoding performance using both the DWT and \architecture\ compression methods applied to the Re90 and Re210 datasets. The compression ratio, $\mathrm{\theta}$, is defined as $\mathrm{S_{\phi_{orig}}}/\mathrm{S_{\phi_{enc}}}$, where $S$ represents a field's size in MB (double precision with three velocity components). $\mathrm{n_\Delta}$ is the downsampling factor applied to the original DNS fields $\mathrm{\phi_{orig}}$, $\mathrm{N_{\phi_{enc}}}$ is the number of points in each spatial dimension of the compressed field $\mathrm{\phi_{enc}}$. ''\architecture`` refers to the size of the \architecture's generator in MB. Note that for $\mathrm{\theta = 2}$ and $\mathrm{\theta = 4}$, only the DWT compression method is employed due to the limitations of SR-GAN three-dimensional upsampling layers.
    }
    \label{tab:table_HIT_results}
\end{table}
The downsampling factor, $\mathrm{n_{\Delta}}$, of the \architecture\ encoding step increases with $\theta$, yet the size of the \architecture\ architecture remains nearly constant. This is because the overall architecture is preserved, with only the upsampling layers being adjusted. Since this modification has a minimal impact on the number of trainable parameters, the changes to model size are negligible. The actual size of each compressed field, $\mathrm{S_{\phi_{enc}}}$ in MB, is reported for each compression ratio, and is consistent regardless of the compression method used. For simplicity, the corresponding number of points in each spatial dimension of the compressed field, $\mathrm{N_{\phi_{enc}}}$, is provided only for the \architecture\ compression method. However, the total number of points aligns with those in $\mathrm{\phi_{enc}}$ when using the DWT method.

\subsection{Encoding/decoding capabilities for various compression ratios}

\begin{figure}[!ht]
    \centering
    \begin{minipage}[b]{0.495\textwidth}
        \centering
        \includegraphics[width=\textwidth]{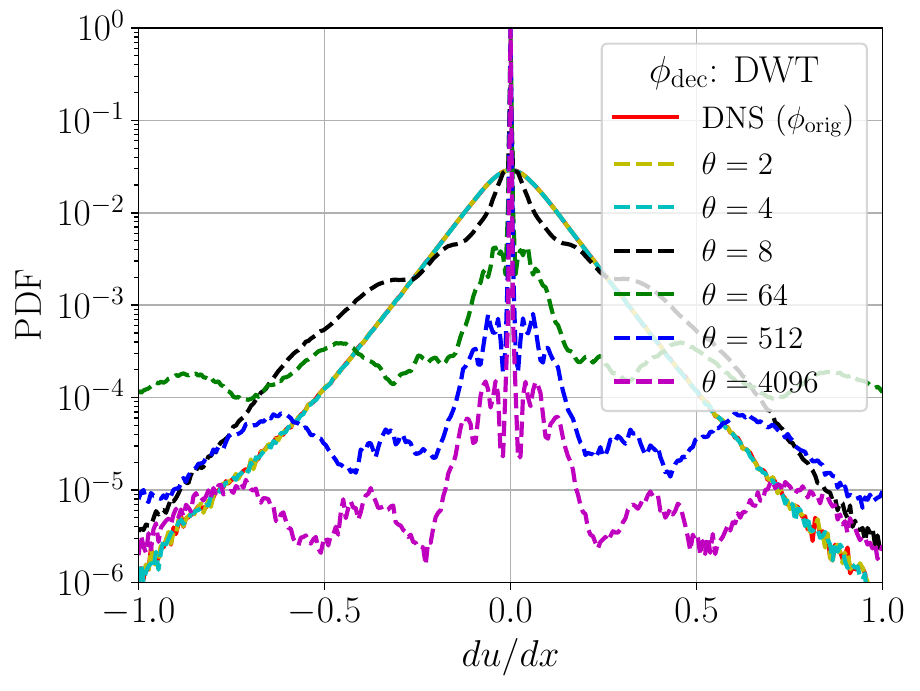}
    \end{minipage}
    \hfill
    \begin{minipage}[b]{0.495\textwidth}
        \centering
        \includegraphics[width=\textwidth]{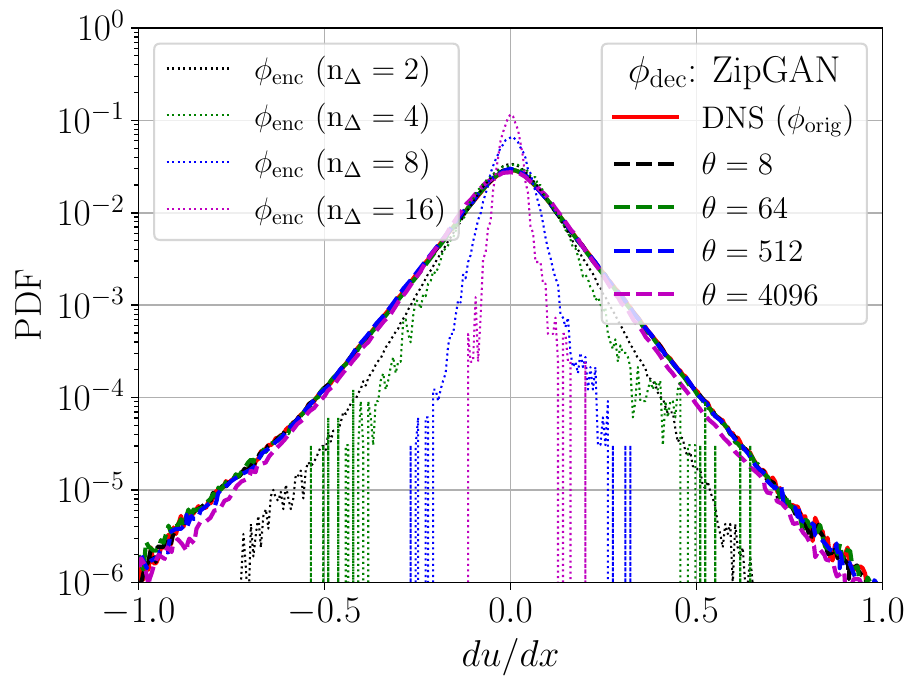}
    \end{minipage}
    \caption{PDF of the normalized velocity gradient of the decoded field, $\mathrm{\phi_{dec}}$, obtained for different compression ratios, $\mathrm{\theta}$, using DWT (left) and \architecture\ (right) compression methods on the Re90 dataset. The input encoded field of the \architecture\ method, $\mathrm{\phi_{enc}}$, is shown for different downsampling factors, $\mathrm{n_{\Delta}}$.}
     \label{fig:PDF_gradient}
\end{figure}

As velocity-gradient statistics are closely associated with fine-scale motion, data compression/decompression methods are assessed based on their ability to reconstruct these gradients. Figure \ref{fig:PDF_gradient} shows the probability density functions (PDFs) of the normalized velocity gradients for $\mathrm{\phi_{enc}}$, $\mathrm{\phi_{dec}}$, and $\mathrm{\phi_{orig}}$, using the DWT (left) and \architecture\ (right) compression methods on the Re90 dataset. For the DWT, significant deviations from the DNS field are observed for $\mathrm{\theta} > 4$, indicating that the quality of the decoded field decreases with $\theta$. In particular, the large distortions observed with the DWT method compared to the original DNS are likely caused by the thresholding process used in wavelet-based compression, which is not based on physical considerations but is necessary to achieve the target compression ratio. This leads to the loss of fine-scale turbulence structures and high-frequency components that are critical for DNS representation. As a result, visible distortions are more prominent, especially in regions with sharp gradients or complex structures. Additionally, due to the local wavelet decomposition employed by the DWT, smaller gradients tend to be predicted more accurately than larger ones, which is generally not observed for \architecture.

\begin{figure}[!ht]
    \centering
    \includegraphics[width=\textwidth]{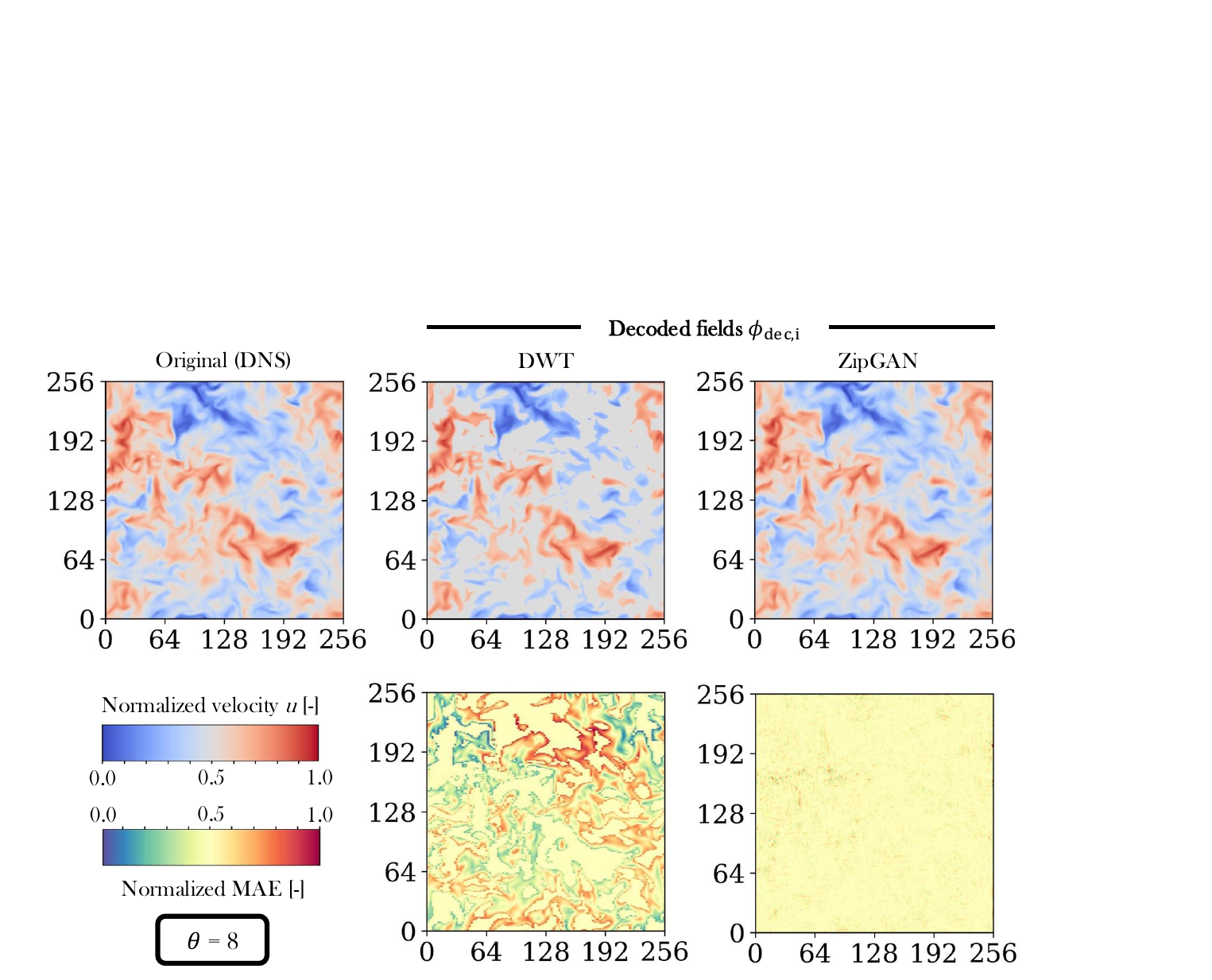}
    \caption{2D slices of instantaneous normalized velocity magnitude and the absolute error of $\phi_{\mathrm{dec,i}}$ fields versus original DNS. The normalized absolute error is given by $\mathrm{\hat{E}=E/\max(E_{\phi_{orig},\phi_{dec,i}},E_{\phi_{orig},\phi_{dec,i}})}$, where $\mathrm{E=\left|\textbf{u}_{\phi_{\mathrm{orig}}} -{\textbf{u}}_{\phi_{\mathrm{dec,i}}}\right|}$ and the subscript $\mathrm{i}$ indicates either the DWT or the \architecture\ data compression methods.}
    \label{fig:contour_plots_8}
\end{figure}

Conversely, the \architecture-decoded fields deviate only marginally from the DNS up to $\mathrm{\theta}=512$. For $\mathrm{\theta}=4096$, \architecture\ slightly underpredicts the large gradients. For an upsampling factor of 2 ($\mathrm{\theta} = 8$), the \architecture\ compression method is already superior to the DWT method. This is quantified in Fig.~\ref{fig:contour_plots_8}, where the centerline slices of the normalized velocity magnitude and its absolute error obtained for both DWT and \architecture\ methods are shown for $\mathrm{\theta}=8$. It is evident that the velocity field decoded by the DWT exhibits large distortions compared to the original DNS data, with the error being especially pronounced for large velocity gradients. On the other hand, the \architecture\ exhibits improved decoding capabilities with only marginal deviations at the smallest scales.

\begin{figure}[!ht]
  \centering
  \includegraphics[width=\textwidth]{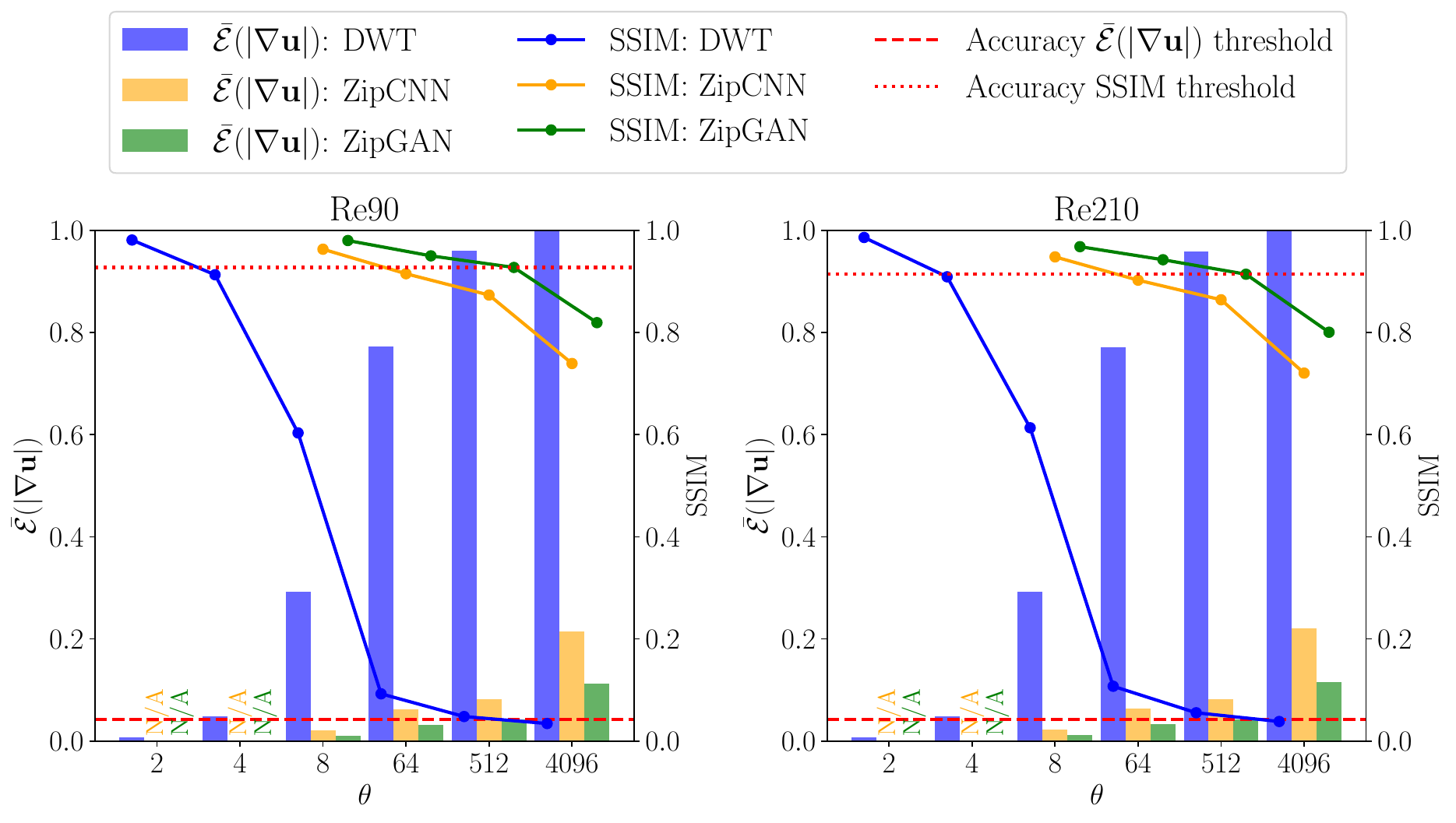}
  \caption{Comparison of the decoding performance at various compression ratios, $\mathrm{\theta}$, based on the averaged NMSE of the velocity gradient, $\bar{\mathcal{E}} ({|\nabla \mathbf{u}|})$, (bars), and the SSIM between $\mathrm{\phi_{dec}}$ and $\mathrm{\phi_{orig}}$ (solid line). The comparison is made using DWT, \cnn, and \architecture\ compression methods on the Re90 dataset (left) and the Re210 dataset (right).}
  \label{fig:Recap_errors}
\end{figure}

The decoding capabilities of each compression method are further analyzed in Fig.~\ref{fig:Recap_errors} (left), where additional metrics including the averaged normalized mean squared error (NMSE) of the velocity gradient $\bar{\mathcal{E}} ({|\nabla \mathbf{u}|})$ (histogram -- lower is better) and the structural similarity index metric (SSIM -- higher is better) between $\mathrm{\phi_{dec}}$ and $\mathrm{\phi_{orig}}$ (solid line) are shown. The NMSE quantifies how well the velocity gradients are preserved after decoding, while the SSIM measures the similarity between $\mathrm{\phi_{dec}}$ and $\mathrm{\phi_{orig}}$ fields. The accuracy thresholds are set based on both the averaged NMSE and SSIM for $\mathrm{\theta = 8}$, and indicate at what $\mathrm{\theta}$ the decoded field starts to deviate from the original, as shown in Fig.~\ref{fig:PDF_gradient} (right). The decoding capabilities are also assessed against a fully supervised SRCNN-based model, called \cnn, introduced in Sec.~\ref{sec:architecture}.

\architecture\ achieves lower NMSEs and higher SSIMs than DWT and \cnn\ methods across all compression ratios. Usage of \cnn\ results in higher NMSEs and lower SSIMs compared to \architecture\ with a degradation factor based on NMSE of roughly 60$\%$, which is still considerably better than the DWT approach. This is due to the preservation of gradient information and structural features by \architecture, particularly at higher compression ratios ($\mathrm{\theta \geq 8}$), as recently demonstrated by Nista \textit{et al.}~\cite{nista2024PRF}. In contrast, DWT performs exceptionally well at lower compression ratios ($\mathrm{\theta \leq 4}$), at which it achieves high accuracy in both NMSE and SSIM metrics. As the compression ratio increases, all methods exhibit increased degradation, with DWT in particular showing severe degradation for $\mathrm{\theta \geq 64}$ while the machine-learning approaches appear to be less sensitive to $\mathrm{\theta}$.

It is important to note that these findings are independent of the Reynolds number of the HIT dataset. In Fig.~\ref{fig:Recap_errors} (right), the same metrics are analyzed when the \architecture\ and \cnn\ architectures, trained solely on the Re90 dataset, are applied to the Re210 dataset. A consistent $\mathrm{dx/\eta}$ ratio is maintained in this analysis to ensure accurate generalization \cite{nista2023proci}. The consistency of the findings suggests that retraining on each new dataset may not be required, provided that $\mathrm{dx/\eta}$ is preserved.

For practical, large-scale applications, \architecture\ provides a significant advantage in data compression. The full-scale Re90 dataset is 95,232 MB in size and can be encoded using \architecture\ to 190 MB ($\mathrm{\theta = 512}$) plus 43 MB for the encoder's state. For similar decoding accuracy, the DWT-encoded dataset requires approximately 19,000 MB, which is approximately $100\times$ more than the \architecture-encoded dataset. A similar compression advantage holds for the larger Re210 dataset, which requires 3,120 GB uncompressed, 6 GB ($\mathrm{\theta = 512}$) for \architecture, and 780 GB for similar decoding accuracy using the DWT.

\subsection{Progressive transfer learning approach}

The superior decoding capabilities of the \architecture\ method can be attributed to (i) the deep architecture of its generator, which efficiently encodes information in the latent space, and (ii) the training process involving adversarial training. The encoding operation, and the filtering and downsampling operation employed to generate LR input fields, are faster than the wavelet-based approaches and do not require training. However, the \architecture\ needs to learn the decoding operation, which is computationally demanding. This training cost must also be considered.

\begin{figure}[!ht]
    \centering
    \includegraphics[width=0.6\textwidth]{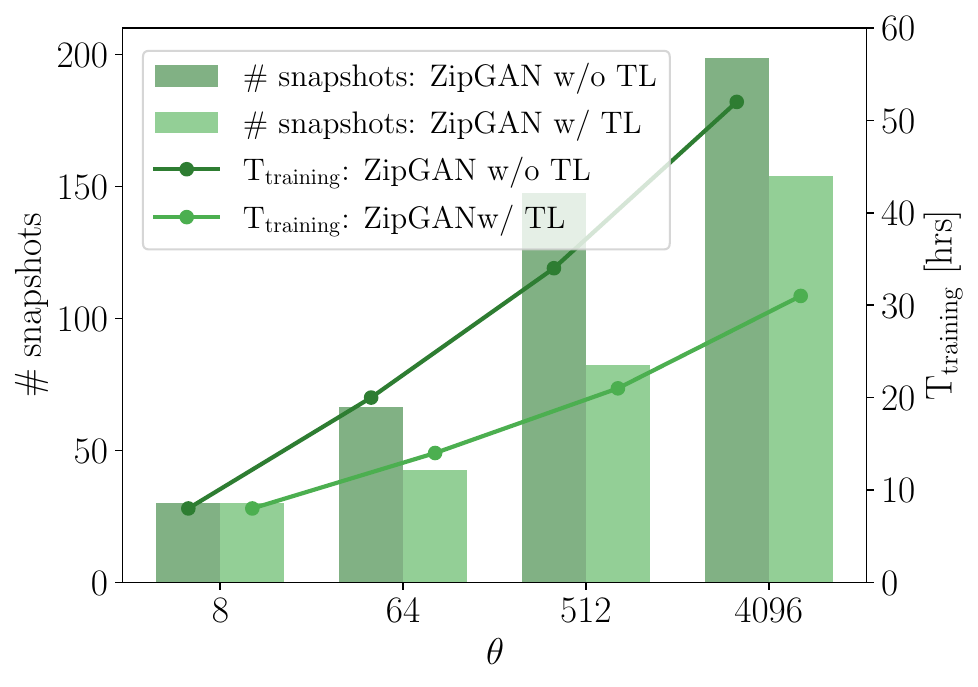}
    \caption{Number of snapshots ($\mathrm{\# \, snapshots}$) required (bars) and training time ($\mathrm{T_{training}}$) (solid line)  needed to achieve the accuracy levels reported in Fig.~\ref{fig:Recap_errors} (left), with and without the use of transfer learning (TL), on the Re90 dataset.}
    \label{fig:Nsnapshots_temporal}
\end{figure}

Figure \ref{fig:Nsnapshots_temporal} (left) highlights the number of snapshots required, and consequently, the time needed for training, to reach the error threshold of $5\%$ reported in Fig.~\ref{fig:Recap_errors} (left). The number of snapshots required and the training time increase sharply with $\theta$ and, consequently, the upsampling factor. This increase occurs because the number of reconstructed features scales with the upsampling factor and increases the complexity and cost of the learning process.

One challenge is the computational expense of saving the original DNS snapshots, and this challenge is compounded by the lengthy training time for SR-GAN models. To mitigate these issues, transfer learning (TL) (see Sec.~\ref{sec:architecture}) is applied for $\mathrm{n_{\Delta}>2}$. TL significantly reduces the number of required snapshots and, as a result, decreases the overall training time, particularly at higher upsampling factors. By leveraging previously acquired knowledge, TL decreases training time to achieve $\bar{\mathcal{E}} ({|\nabla \mathbf{u}|})<0.05$ by approximately $50\%$ with every doubling of $\mathrm{n_{\Delta}}$. TL is clearly effective in enhancing training efficiency and reducing computational demands.

\subsection{Enhancing temporal resolution}

\begin{figure}
    \centering
    \begin{minipage}[c]{0.495\textwidth}
        \centering
        \includegraphics[width=\textwidth]{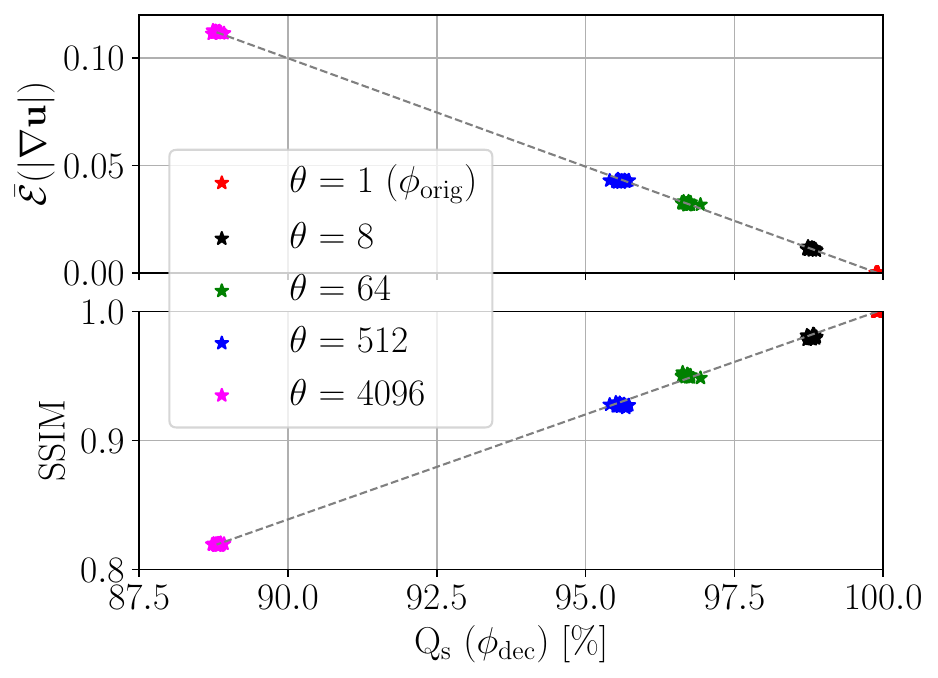}
    \end{minipage}
    \hfill
    \begin{minipage}[c]{0.495\textwidth}
        \centering
        \includegraphics[width=\textwidth]{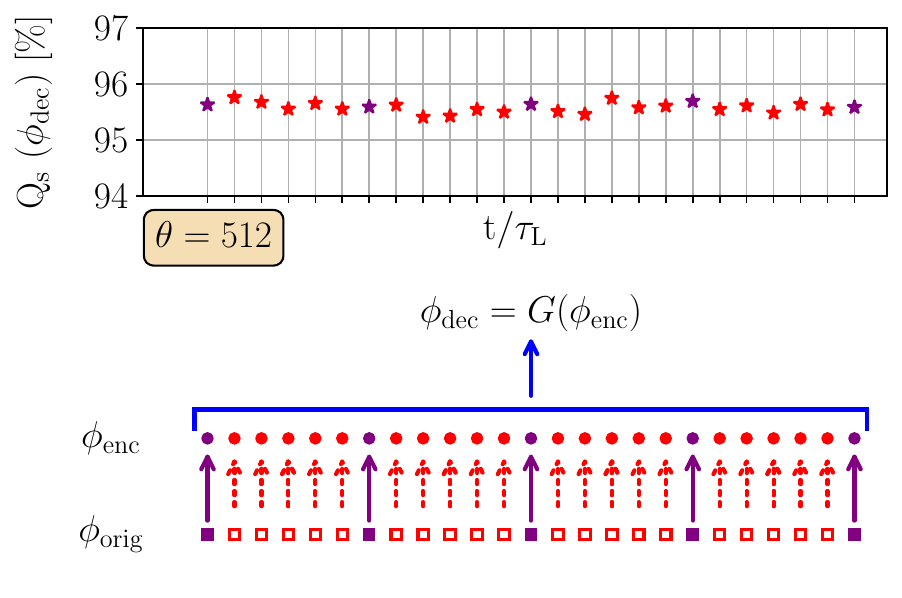} 
    \end{minipage}
    \caption{Left: Correlation of the quality score, $\mathrm{Q_s}$, with both the NSME of the velocity gradients, $\bar{\mathcal{E}} ({|\nabla \mathbf{u}|})$ (top frame), and the SSIM (bottom frame) obtained from the $\phi_\mathrm{dec}$ using \architecture\ across different compression ratios, $\theta$. Right: Qualitative illustration of the SR-GAN-based approach for enhancing temporal resolution. Original DNS fields ($\mathrm{\phi_{orig}}$, purple squares) are saved at intervals, filtered, and downsampled (purple arrows) to create in-sample encoded fields ($\mathrm{\phi_{enc}}$, purple circles) for SR-GAN training. Out-of-sample fields are only saved as encoded fields ($\mathrm{\phi_{enc}}$, red circles) and are decoded by the SR-GAN to enhance temporal resolution. The pretrained discriminator provides a quality score ($\mathrm{Q_{s}}$) which is used to assess both in-sample (purple asterisks) and out-of-sample (red asterisks) decoded fields.}
    \label{fig:Temporal_discretization}
\end{figure}

Compared to standard supervised SR-CNN approaches, SR-GAN-based training is costlier due to the discriminator network. 
However, the discriminator is useful for more than distinguishing between $\mathrm{\phi_{dec}}$ and $\mathrm{\phi_{orig}}$ fields during the adversarial training. Since the pretrained discriminator contains DNS physical information in its latent space, it can provide a ``quality score'' of the $\mathrm{\phi_{dec}}$ field,
\begin{equation}
\mathrm{Q_s} = \mathbb{E}_{\phi_{\ENCsub} \sim p_\mathrm{enc}}[\log(\sigma(D(G(\mathrm{\phi_{enc}}))))],
\label{eqn:quality_score}
\end{equation} 
where higher values of $\mathrm{Q_s}$ indicate greater confidence in the authenticity of the decoded field. To assess the reliability of the $\mathrm{Q_s}$ metric and ensure that the discriminator loss accurately reflects the $\phi_{\mathrm{dec}}$ quality, calibration of the discriminator is validated in Fig.~\ref{fig:Temporal_discretization} (left). Here, the $\mathrm{Q_s}$ obtained from each available $\phi_{\mathrm{dec}}$ field is correlated with both $\bar{\mathcal{E}} ({|\nabla \mathbf{u}|})$ and SSIM for each $\theta$. A nearly perfect positive correlation between $\mathrm{Q_s}$ and SSIM is obtained, indicating that higher $\mathrm{Q_s}$ values align with higher structural similarity of the decoded fields. Similarly, $\mathrm{Q_s}$ exhibits a negative correlation with NSME, suggesting that lower NSME values correspond to higher values of $\mathrm{Q_s}$. These consistent trends across all the $\theta$ values highlight the robustness of $\mathrm{Q_s}$ as a metric for quantifying decoded field quality. Moreover, effective calibration of the discriminator is demonstrated, as it accurately identifies $\mathrm{\phi_{orig}}$ fields with a quality score of 99.9$\%$. The $\mathrm{Q_s}$ obtained on the $\phi_\mathrm{dec}$ fields using both the DWT and \architecture\ are reported in Tab.~\ref{tab:quality_score}, corroborating our previous analysis reported in Fig.~\ref{fig:Recap_errors} (left).
\begin{table}[!ht]
    \centering
    \resizebox{\textwidth}{!}{%
    \begin{tabular}{c|c|c|c|c|c|c|c}
    \toprule 
    $\mathrm{\theta}$ & 1 & 2 & 4 & 8 & 64 & 512 & 4096 \\
    \midrule
    \makecell{Compression \\ method}  & Orig. & DWT & DWT & \makecell[l]{DWT \, \architecture} & \makecell[l]{DWT \, \architecture} & \makecell{DWT \, \architecture} & \makecell[l]{DWT \, \architecture} \\
    \hline
    \makecell{$\mathrm{Q_{s}}$  $\mathrm{[\%]}$ -- Re90}  & 99.9 & 99.7 & 99.2 & \makecell[l]{\, 36.6  \quad \, 98.8} & \makecell[l]{\, 9.3 \quad \, \, 96.7} & \makecell[l]{\, 1.6 \quad \, \, 95.6} & \makecell[l]{\, 0.8 \quad \, \, 88.8} \\
    \bottomrule
    \end{tabular}
    }
    \caption{Quality score, $\mathrm{Q_{s}}$, of the $\phi_\mathrm{dec}$ fields using both the DWT and \architecture\ compression methods applied to the Re90 dataset. Higher $\mathrm{Q_{s}}$ values indicate greater confidence.}
    \label{tab:quality_score}
\end{table}

Turbulent flow applications additionally require high temporal resolution for post-processing and analysis, but storing DNS fields at every time step is impractical due to memory and computational constraints. Figure \ref{fig:Temporal_discretization} (right) qualitatively illustrates the ability of the SR-GAN method to address this limitation. The DNS fields ($\mathrm{\phi_{orig}}$, purple squares) are saved at infrequent time intervals, are filtered and downsampled (purple arrows) to produce encoded fields $\mathrm{\phi_{enc}}$ (purple circles), which are used during SR-GAN training and are referred to as ``in-sample.'' These fields serve as the basis for the training, validation, and testing datasets since the corresponding DNS fields are available. Rather than storing DNS fields at high temporal frequencies (red empty squares), our strategy instead stores filtered and downsampled $\mathrm{\phi_{enc}}$ fields between DNS snapshots (red circles). These out-of-sample $\mathrm{\phi_{enc}}$ fields can be decoded by the SR-GAN model after training to enhance the temporal resolution of the dataset with additional $\mathrm{\phi_{dec}}$ fields.

In this regard, a unique feature of SR-GAN-based compression is its ability to use the pretrained discriminator to evaluate the accuracy of decoded out-of-sample $\mathrm{\phi_{dec}}$ fields, even in the absence of corresponding DNS ($\mathrm{\phi_{orig}}$) data. As depicted in Fig.~\ref{fig:Temporal_discretization} (right), the discriminator outputs the quality score, $\mathrm{Q_{s}}$, for both in-sample (purple asterisks) and out-of-sample (red asterisks) decoded fields, $\mathrm{Q_s}(\mathrm{\phi_{dec}})$. As expected, higher $\mathrm{Q_s}$ values are observed for in-sample $\mathrm{\phi_{enc}}$ fields compared to the out-of-sample $\mathrm{\phi_{enc}}$ fields, as the model was explicitly trained on the former. Despite this, the out-of-sample $\mathrm{\phi_{enc}}$ fields demonstrate sufficient accuracy, making them viable for enhancing the temporal resolution of the DNS dataset.

\subsection{Decoding time}

The decoding time, $\mathrm{T_{\Gamma}}$, is quantified to provide a more complete understanding of the tradeoffs between the two compression methods evaluated. As shown in Table \ref{tab:decoding_time}, the $\mathrm{T_{\Gamma}}$ for DWT is insensitive to $\theta$. This is because the decoding process scales linearly with the DNS size and is minimally influenced by the compression ratio. Conversely, the decoding time for the \architecture\ method increases with $\theta$, as increasing $\theta$ requires an increasing number of operations to reconstruct the full-resolution data. The \architecture\ framework’s ability to leverage GPU acceleration enables it to maintain a decoding-time advantage over the DWT, which we have tested for CPU-only calculations, even at very high compression ratios.

\begin{table}[!ht]
       \centering
    \begin{tabular}{c|c|c|c|c|c|c}
    \toprule 
    $\mathrm{\theta}$ & 2 & 4 & 8 & 64 & 512 & 4096 \\
    \midrule
    \makecell{Compression \\ method}  & DWT & DWT & \makecell[l]{DWT \, \architecture} & \makecell[l]{DWT \, \architecture} & \makecell{DWT \, \architecture} & \makecell[l]{DWT \,  \architecture} \\
    \hline
    \makecell[l]{$T_{\Gamma}$ [s] -- Re90}  & 0.29 & 0.29 & \makecell[l]{\, 0.31  \quad 0.013} & \makecell[l]{\, 0.32   \quad 0.062} & \makecell[l]{\, 0.34 \quad 0.28} & \makecell[l]{\, 0.34 \quad  1.63} \\
    \makecell[l]{$T_{\Gamma}$ [s] -- Re210} & 5.84 & 5.87 & \makecell[l]{\, 5.88  \quad 0.038} & \makecell[l]{\, 5.89   \quad 0.35} & \makecell[l]{\, 5.91 \quad 1.76} & \makecell[l]{\, 5.92 \quad 13.74} \\
    \bottomrule
    \end{tabular}
    \caption{Comparison of the decoding time, $\mathrm{T_{\Gamma}}$, using both the DWT and \architecture\ compression methods applied to the Re90 and Re210 datasets for different compression ratios. The DWT employs 48 CPUs and \architecture\ uses a single GPU.}
    \label{tab:decoding_time}
\end{table}



\section{Conclusion}
\label{sec:conclusion}

We investigate the performance of \architecture, a SR-GAN-based compression and decompression method, for the storage and transfer of DNS HIT datasets and compare it with the performance of the DWT method. The \architecture\ architecture, with its generator acting as a decoding operator, accurately reconstructs high-resolution fields from their encoded, low-resolution counterparts. Once trained, the \architecture's generator can decode fields solely from the encoded input, i.e., without requiring further access to the original DNS data.

The \architecture-based method maintains high fidelity even at a compression ratio of $\theta=512$, accurately preserving small-scale features and turbulence intermittency, while considerable encoding/decoding error is introduced by the DWT for $\theta\geq 4$. While marginal degradation of \architecture-reconstructed fields is observed for $\theta=4096$, the \architecture's overall performance exceeds that of the DWT for $\theta\geq8$. This suggests that multiscale SR-GAN frameworks for data compression can significantly improve compression ratio and reconstruction accuracy, with adversarial training being essential for achieving near-lossless decoding. These findings appear to be independent of the Reynolds number of the HIT dataset, provided that $\mathrm{dx}/\eta$ is preserved between training and testing datasets.

Transfer learning is employed to address the scalability challenges associated with computationally complex SR-GAN frameworks. This significantly decreases the number of snapshots required for training and also reduces training time. The combination makes \architecture\ more practical and efficient, particularly for high compression ratios. 
 
Additionally, the \architecture\ framework is used to enhance the temporal resolution of datasets without requiring additional simulation time. By generating high-quality intermediates of the decoded fields from encoded snapshots, the \architecture\ is able to fill the temporal gaps between stored DNS fields. This is enabled by the discriminator network, which provides a reliable quality score for reconstructed data even in the absence of the original DNS data. Thus, the discriminator network serves as a built-in quality-control mechanism unique to the \architecture\ architecture. This ensures that only high-fidelity decoded fields are retained and further enhances the utility of SR-GAN-based architectures for large-scale DNS data compression.

In summary, \architecture\ offers a highly efficient and scalable solution for compressing large DNS datasets of turbulent flows without introducing large decoding errors at high compression ratios. It is particularly well-suited for applications requiring the preservation of fine-scale features and sufficient temporal resolution, such as three-dimensional turbulent flow datasets.

In order to enhance the reproducibility of this study, provide clarity on technical aspects, and facilitate faster development, access to our GIT repository will be granted upon request.


\begin{acknowledgement}
The research leading to these results has received funding from the German Federal Ministry of Education and Research (BMBF) and the state of North Rhine-Westphalia for supporting this work as part of the NHR funding and the European Union under the European Research Council Advanced Grant HYDROGENATE, Grant Agreement No.~\textit{101054894}. The authors gratefully acknowledge the computing resources granted by the NHR4CES Resource Allocation Board on the high-performance computer CLAIX at the NHR Center RWTH Aachen University, and F.~Orland and Dr.~T.~Gerrits for their exceptional support and contributions to this research project.
\end{acknowledgement}




\bibliographystyle{ieeetr}
\bibliography{references}

\end{document}